\documentclass{aastex}
\usepackage{spr-astr-addons}
\usepackage{url}

\newcommand{\emaila}{goicol@unican.es}

\begin{document}

\title{Gravitationally lensed QSOs in the ISSIS/WSO-UV era}
%% Running heads
\shorttitle{GLQs in the ISSIS/WSO-UV era}
\shortauthors{Goicoechea et al.}

\author{L. J. Goicoechea}
\affil{Departamento de F\'\i sica Moderna, Universidad de Cantabria, 
Avda. de Los Castros s/n, 39005 Santander, Spain}
\email{\emaila} 
\and 
\author{V. N. Shalyapin}
\affil{Institute for Radiophysics and Electronics, National Academy 
of Sciences of Ukraine, 12 Proskura St., 61085 Kharkov, Ukraine}
\and
\author{R. Gil-Merino}
\affil{Departamento de F\'\i sica Moderna, Universidad de Cantabria, 
Avda. de Los Castros s/n, 39005 Santander, Spain}

\begin{abstract}
Gravitationally lensed QSOs (GLQs) at 1 $\leq z \leq$ 2 play a key role in 
understanding the cosmic evolution of the innermost parts of active galaxies
(black holes, accretion disks, coronas and internal jets), as well as the  
structure of galaxies at intermediate redshifts. With respect to studies of 
normal QSOs, GLQ programmes have several advantages. For example, a 
monitoring of GLQs may lead to unambiguous detections of intrinsic and 
extrinsic variations. Both kinds of variations can be used to discuss 
central engines in distant QSOs, and mass distributions and compositions of 
lensing galaxies. In this context, UV data are of particular interest, since
they correspond to emissions from the immediate surroundings of the 
supermassive black hole. We describe some observation strategies to analyse 
optically bright GLQs at z $\sim$ 1.5, using ISSIS (CfS) on board World Space 
Observatory-Ultraviolet.
\end{abstract}


%\section*{}
%\label{sec:intro}
\section{Gravitationally lensed QSOs in the UV: central engines}

Clusters, groups and individual galaxies are so massive and compact that light 
rays from background sources are gravitationally deflected by them. In the strong 
lensing regime, the involved gravitational fields bend light to form several 
images of the same background object \citep{blan92,schn92,wamb98,schn06}. The 
first gravitationally lensed QSO (GLQ) was discovered by \citet{wals79} 30 years 
ago, and there are currently more than 100 known GLQs. Most of them are active 
galactic nuclei (AGN) at $z >$ 1 \citep{cast10,sqls10}. 

\begin{figure}[t]
\includegraphics[angle=-90,width=\columnwidth]{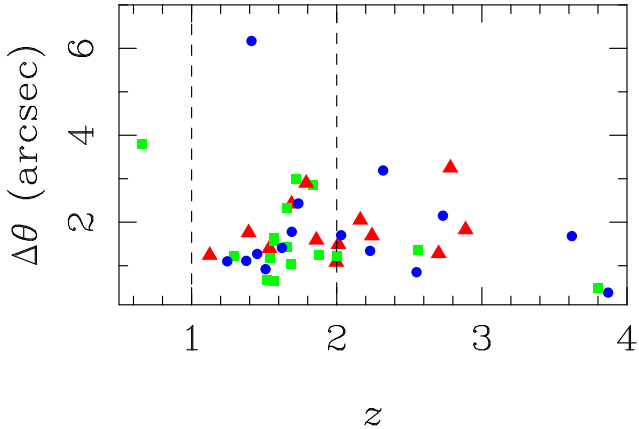}
\caption{CASTLES + SQLS sample of bright GLQs. We display angular sizes and 
redshifts for GLQs having at least two images with $V <$ 20 mag. Taking the 
$V$-band flux of the brightest image of each GLQ ($V_{\star}$) into account, the
sample is divided in three different subsamples: $V_{\star} <$ 17.5 mag (blue 
circles), 17.5 $< V_{\star} <$ 18.5 mag (green squares), and $V_{\star} >$ 18.5 
mag (red triangles). Two vertical lines define the region of interest (1 $< z <$ 
2)}
\label{SelectionV20}
\end{figure}

We initially consider the population of optically bright GLQs. From the 
CASTLES\footnote{\url{http://www.cfa.harvard.edu/castles/}} and 
SQLS\footnote{\url{http://www-utap.phys.s.u-tokyo.ac.jp/~sdss/sqls/}} databases, 
we select 46 GLQs showing at least two images with $V <$ 20 mag. This GLQ sample 
is depicted in Fig.~\ref{SelectionV20} (only two sample members with angular size 
$\Delta \theta > 10\arcsec$ are not included in the plot, where $\Delta \theta$ 
represents the maximum separation of any pair of images, twice the average 
distance of images from the lens center or something like that). We also classify 
the population in terms of the $V$-band flux of the brightest image of each GLQ 
($V_{\star}$): $V_{\star} <$ 17.5 mag (blue circles), 17.5 $< V_{\star} <$ 18.5 
mag (green squares), and $V_{\star} >$ 18.5 mag (red triangles). About 50\% of 
the GLQs with $V_{\star} <$ 17.5 mag have their redshifts in the interval 1 $< z 
<$ 2 (see the region between both vertical dashed lines in 
Fig.~\ref{SelectionV20}), and we focus on this densely populated cosmic shell 
containing 61\% (28 out of 46) of the whole CASTLES + SQLS sample, i.e., the 
population of bright GLQs at $z \sim$ 1.5.

%----------------------------------------------------------- SelectionV20
%______________________________________________________________

A lensing galaxy (or group/cluster) acts as a natural telescope, so it produces a 
GLQ with magnified images (with respect to the single image of a non-lensed QSO 
at the same $z$ and having the same intrinsic luminosity). However, the dust in 
the galaxy may significantly demagnify some GLQ images at certain wavelengths 
\citet{falc99}. Moreover, the presence of high column densities of neutral 
hydrogen (in intervening objects at $z \leq$ 1.5) could greatly hinder or make 
impossible the detection of photons at $\lambda_{obs} <$ 2280 \AA, i.e., $\lambda 
<$ 912 \AA\ (= Lyman limit) in the GLQ rest frame ($z \sim$ 1.5). At the redshift 
of interest, we remark that optical observations in the $BVR$ Bessell and $gr$ 
SDSS bands correspond to far/middle UV (FUV/MUV) emission ($\lambda \sim$ 
1700$-$2600 \AA). Hence, even in the worst case (total absorption below the Lyman 
limit), UV observations at $\lambda_{obs} \sim$ 2300-4000 \AA\ (EUV/FUV emission:
$\lambda \sim$ 920-1600 \AA) are possible. These have an extraordinary relevance 
in the study of the innermost accretion flow in a distant QSO, and thus, to 
reveal its hottest and most compact sources around the supermassive black hole 
\citep[SMBH;][]{pete09}. 

\section{Time domain studies of optically bright GLQs at $z \sim$ 1.5}

\subsection{Intrinsic variability}

A monitoring campaign of a normal (non-lensed) QSO at $z \sim$ 1.5 produces light 
curves in the UV/optical continuum whose variations are difficult to interpret: 
intrinsic variability or fluctuations caused by the intervening medium? However, 
the presence of a gravitational deflector at intermediate redshift may result in 
several (usually two or four) images of the same QSO. As the ray paths are 
different for different images, the corresponding magnifications and traveltimes 
do not agree with each other. Thus, in the absence of significant extrinsic 
variations (see the opposite case below), each intrinsic flare is observed in two 
or more images with different arrival times. This allows observers to fairly 
interpret the nature of the variability, and even more, the light curves of the 
images can be combined to make one better-sampled intrinsic record 
\citep[e.g.][]{kawa98,goic08a}. 

A structure function analysis is a powerful technique to quantify typical 
luminosity variabilities at different rest frame lags and to check for timescales 
of intrinsic flares \citep[e.g.][]{cidf00,coll01}. For two optically bright GLQs 
at $z$ = 1.4 (SBS 0909+532 and Q0957+561), we recently reported the possible 
existence of time-symmetric flares (with symmetric rise and decay) at $\lambda 
\sim$ 1900-2600 \AA\ and lasting $\sim$ 100 d \citep{goic10a}. As both GLQs have 
similar SMBH mass \citep{peng06}, the symmetry and timescale could be related to 
this physical quantity. There is also evidence that 170-d asymmetric flares are 
generated in the nuclear region of Q0957+561, but these events seem to have an 
intermittent character \citep{goic08b}. Future studies involving larger samples 
and longer observation periods will confirm/reject the possible correlation 
between symmetry/timescale and SMBH mass, as well as the generation of asymmetric 
flares in an intermittent way in some GLQs. The radio-loud GLQ Q0957+561 also has 
an achromatic structure function at $\lambda \sim$ 1900-2600 \AA, suggesting the
existence of an illuminated accretion disk \citep[reprocessing 
scenario;][]{coll91,krol91}: high-energy (X-ray/EUV) flares from a source above 
the SMBH are reprocessed by disk rings into lower energy events. 

For a given AGN, time delays between variations at different UV/optical continuum 
wavelengths constrain the SMBH accretion physics. The standard reprocessing 
scenario consists of a very compact X-ray source located at a height of a few 
Schwarzschild radii ($h \sim$ a few $R_{Schw}$) above the SMBH, which illuminates 
a standard geometrically-thin and optically-thick accretion disk \citep{shak73}. 
This corona-disk coupling scenario predicts light travel time delays between an 
X-ray central flare and its replicas (reprocessed versions) in different rings of 
the accretion disk. The smallest radii correspond to the hottest regions and 
shortest wavelength emissions, so light travel time delays $\tau = R/c$ ($R > h$ 
is an arbitrary radius and $c$ is the speed of light) can be rewritten as 
wavelength-dependent lags $\tau \propto \lambda^{4/3}$ 
\citep[e.g.][]{coll99,serg05,cack07}. Light curves of many local AGN show 
UV/optical inter-wavelength lags supporting the standard reprocessing scenario 
\citep[e.g.][]{serg05}. However, there are also indications that UV/optical 
fluctuations in several local AGN cannot be produced from standard reprocessing 
of X-ray flares associated with observed X-ray variations. 

For example, \citet{arev08} simulated thermally-reprocessed $B$-band light curves 
of the local QSO MR 2251-178, using the technique by \citet{kaza01} and the 
observed X-ray record as input. They showed that even observed short-timescale 
(lasting tens of days) B-band fluctuations do not agree with standard 
reprocessing simulations. If we concentrate on the two consecutive fluctuations 
in the 3950-4100 (MJD-50000) period \citep[see Fig. 9 in][]{arev08}, only the 
green line seems to account for the B-band data. This line comes from a 
non-standard reprocessing simulation with $h$ = 50 $R_{Schw}$, i.e., the 
irradiating source does not correspond to a X-ray corona just above the SMBH. 
Although the X-ray source might be really located at the base of a jet, we note 
that at least one best-fit parameter, the observer latitude, is physically 
unplausible (an edge-on disk generates zero flux at the observer and it is 
inconsistent with a type 1 AGN). \citet{arev09} also showed that the reprocessor 
of rapid X-ray flares in the Seyfert nucleus NGC 3783 could correspond to the 
broad emission-line region (instead the accretion disk). In brief, 
short-timescale UV/optical variations are very likely related to reprocessing of 
high-energy events, but there is no standard scenario working in all local AGN. 
The main high-energy source may not be the X-ray corona (X rays from the base of 
a jet, EUV emission from a non-flat innermost region of the accretion flow, etc) 
and/or the reprocessor may be different than a standard accretion disk 
(non-standard disk, broad emission-line region, dust torus, etc). 
 
\begin{figure}[t]
\includegraphics[angle=0,width=\columnwidth]{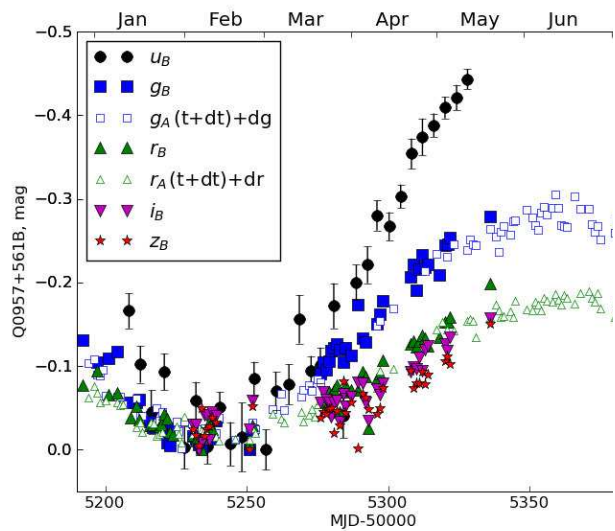}
\caption{Preliminary light curves of Q0957+561B over the first semester of 2010. 
Records of Q0957+561A (but 14 months before) are also depicted for comparison 
purposes (open symbols; see main text). The UV/optical/NIR variability of the B 
image of the distant GLQ has been monitorized with Swift/UVOT (black circles; the 
rest frame wavelength is $\lambda \sim$ 1440 \AA), and the Liverpool Robotic 
Telescope: $\lambda \sim$ 1940 \AA\ (filled blue squares), $\lambda \sim$ 2560 
\AA\ (filled geen triangles), $\lambda \sim$ 3100 \AA\ (purple inverted 
triangles), and $\lambda \sim$ 3710 \AA\ (red stars). The light curves are 
shifted in magnitude to overlap them at day 5250}
\label{Montwin}
\end{figure}

Time lags between variations at different wavelengths has been well studied only
for local AGN. However, very recently we have detected a prominent optical 
variation in the image A of the GLQ Q0957+561, and searched for the imprints of
a hypothetical intrinsic flare in the light curves of the B image, but 14 months
later \citep[the well-known time delay between both images, e.g.][]{shal08}. 
Fortunately, we have observed the GLQ ($\Delta \theta \sim$ 6$\arcsec$) with 
several facilities at several 
wavelengths: X-rays (Chandra), NUV (Swift/UVOT) and optical/NIR (Liverpool 
Robotic Telescope). In Fig.~\ref{Montwin}, we show preliminary NUV/optical/NIR 
light curves of Q0957+561B (until mid-May 2010) from our monitoring campaign with 
the UVOT on board Swift (black circles) and the Liverpool Robotic Telescope (rest 
of filled symbols). Open symbols trace the time evolution of Q0957+561A (14 
months before) and demonstrate the intrinsic origin of the variability (e.g. the
filled blue squares correspond to Q0957+561B data in the $g$ SDSS band, and these
are consistent with the open blue squares describing the behaviour of Q0957+561A
in the same optical filter). Our NUV/optical/NIR follow-up covers the rest-frame 
spectral region $\lambda \sim$ 1440-3700 \AA, so we have lost EUV emissions at 
$\lambda \sim$ 920-1200 \AA, which probably come from the innermost accretion 
flow. In any case, the simple use of the $U$ filter in the UVOT filter wheel 
allows us to properly analyse the FUV emission at $\lambda \sim$ 1440 (8.6 eV). 
This is also expected to be created in internal rings of the accretion disk 
around the SMBH. At present, we are comparing the flux fluctuations of 
Q0957+561B from X-rays to NIR waves.

%----------------------------------------------------------- Montwin
%______________________________________________________________

Time-delayed responses of emission lines to continuum variations can be used to 
infer sizes of broad emission-line regions, while FWHMs of emission lines are 
directly related to the kinematics of line-emitting clouds. Both the geometry and 
kinematics of broad emission-line regions are useful tools to estimate masses of 
central SMBHs \citep[e.g.][]{pete09}. Unfortunately, there are no spectral 
monitoring campaigns of GLQs (with the exception of COSMOGRAIL's recent efforts 
to follow-up the spectral variability of Q2237+0305 - see below, and an ongoing 
programme to take spectra of the two images of Q0957+561 on a regular basis), 
which might be much more efficients than programmes with non-lensed QSOs. This is
because effective sampling periods are twice (double QSO) or even four times 
(quadruple QSO).

Apart from unambiguous detections of intrinsic fluctuations and important 
improvements in effective samplings, it is clear that studies of GLQs have much 
more potential than those from non-lensed QSOs. Thus, time delays between lensed 
images are related to lensing mass distributions, and provide extremely valuable 
information on dark matter halos around galaxies at intermediate redshifts 
\citep[e.g.][]{schn06}. From very distant GLQs, one can also infer unknown 
redshifts (not only dark halos) of lensing objects. For example, the time delays 
in the quadruple QSO H1413+117 (Cloverleaf) at $z$ = 2.56 have been used to 
refine the lens model and to estimate the previously unknown lens redshift 
\citep{goic10b}.

\subsection{Extrinsic variability}

%----------------------------------------------------------- Imagecross
\begin{figure}[t]
\includegraphics[angle=0,width=\columnwidth]{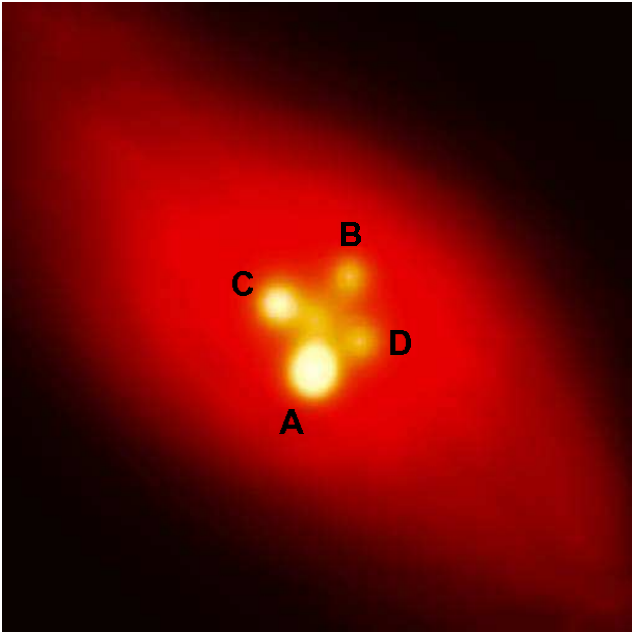}
\caption{Einstein Cross at optical wavelengths. Apart from the four images 
A-D of the distant GLQ Q2237+0305, one can also see the local face-on spiral 
galaxy G2237+0305 acting as a gravitational lens}
\label{Imagecross}
\end{figure}
%______________________________________________________________
%----------------------------------------------------------- Glitpcross
\begin{figure}[t]
\includegraphics[angle=-90,width=\columnwidth]{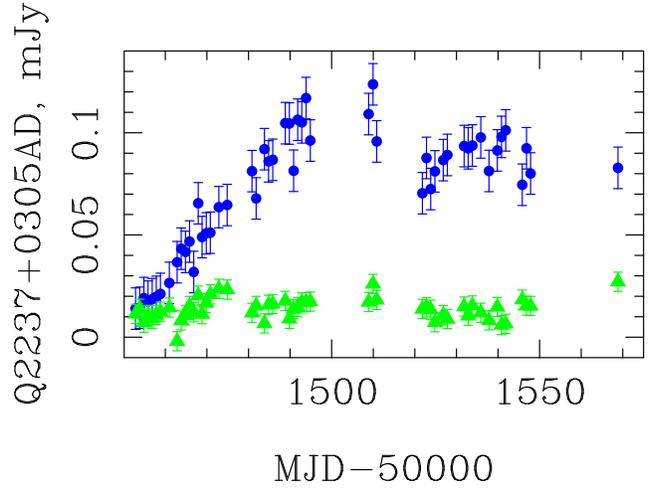}
\caption{GLITP $V$-band fluxes (mJy) of Q2237+0305A (blue circles) and 
Q2237+0305D (green triangles) between October 1999 and February 2000. The 
observations were made with the 2.56-m Nordic Optical Telescope, and the two 
light curves are shifted in flux to overlap them at day 1450}
\label{Glitpcross}
\end{figure}
%______________________________________________________________

Apart from the constant macrolens magnification (due to the lensing structure as 
a whole) of an image of a GLQ, an additional variable microlens magnification may 
be generated by compact objects (moving stars or MACHOs) in the main lensing 
galaxy \citep[gravitational microlensing; e.g.][]{schn06}. In principle, moving 
clouds of dust may also produce a variable transmission \citep[e.g.][]{goic05}, 
and thus, some extrinsic variability. However, most studies focused on the 
microlensing interpretation of extrinsic signals in GLQs. When the extrinsic 
variability can be unambiguously separated from the intrinsic one 
\citep[e.g.][]{para06}, resulting extrinsic fluctuations may be related to the 
nature of the lensed and microlensed source, as well as the composition of the 
main lensing galaxy \citep[e.g.][]{shal02,koch04,para06}.

Microlensing (extrinsic) variability was initially discovered in the GLQ 
Q2237+0305 \citep[Einstein Cross;][]{irwi89}, and we mainly comment results on 
this emblematic QSO. The gravitational lens system is formed by a distant QSO at 
$z$ = 1.695 and a local lens galaxy (face-on spiral at $z$ =0.039) that 
quadruples the images of the QSO  \citep{huch85}. This compact lens system with
$\Delta \theta \sim$ 2$\arcsec$ appears in Fig.~\ref{Imagecross}. Some optical 
continuum 
monitoring campaigns have led to very accurate and reasonably well-sampled light 
curves of the four images of Q2237+0305, e.g. the OGLE \citep{wozn00}, GLITP 
\citep{alca02} and COSMOGRAIL \citep{eige08} records. The OGLE $V$-band records 
cover a time frame of about 10 years, while the GLITP $VR$-band curves only last 
a few months but with a daily sampling, bad weather or technical problems aside
\citep[see also the Maidanak light curves in the $R$ band;][]{vaku06}. The 
COSMOGRAIL monitoring is the only spectroscopic follow-up ($\lambda \geq$ 1450 
\AA\ in the rest frame of the QSO) of a GLQ to date. In Fig.~\ref{Glitpcross}, 
the GLITP $V$-band fluxes (mJy) of Q2237+0305A and Q2237+0305D are plotted 
together as blue circles and green triangles, respectively. The time delay 
between the images A and D is less than 2 days \citep{vaku06}, so the time-axis
in Fig.~\ref{Glitpcross} practically traces the emission time. There is a
significant fluctuation in A, which is not detected in D at similar emission 
times. This extrinsic fluctuation allowed us to discuss the $V$-band source 
intensity profile and size, the mass of the SMBH, the mass accretion 
rate, the lens transverse velocity, and the typical mass of stars in the bulge
of the local spiral galaxy \citep{shal02,goic03,gilm06}. 

The timescale of microlensing variability depends on the redshift of (distance 
to) the main lensing galaxy: more distant lenses produce longer variations. 
Therefore, the Einstein Cross is of particular interest because of the 
observed short-timescale microlensing fluctuations (e.g. Fig.~\ref{Glitpcross}).
For non-local lenses, one observes long-term microlensing effects that induce 
small (optical) flux variations on timescales of several months 
\citep[e.g.][]{gayn05,fohl07,shal09}. QSO emission regions have different sizes 
in different wavelengths, and microlensing effects are stronger for more compact
sources. For example, \citet{pool07} used observed X-ray and optical flux ratios 
between images of ten quadruple GLQs to show that X-ray sources (hot coronas 
above SMBHs?; see above) are much more strongly microlensed (more compact) than 
sources associated with optical observations (accretion disks around SMBHs are
prime candidates). Thus, the UV extrinsic variability should be stronger than the
extrinsic variability at optical wavelengths, and it should probe innermost 
regions of accretion disks.

\section{GLQ programmes using ISSIS (CfS) on board World Space 
Observatory-Ultraviolet (WSO-UV)}

The aim of the WSO-UV mission is to study the Universe in the UV spectral range 
\citep{shus09}, but also including complementary optical/NIR capabilities 
(4000-10000 \AA). This 1.70-m space telescope will be launched in 2014, and it 
will be the replacement for the Hubble Space Telescope (HST) in many ways. While 
the 4000-10000 \AA\ interval is not a problem for ground-based instruments, the 
1000-4000 \AA\ range is problematic or beyond the sensitivity of ground-based 
telescopes. Therefore, WSO-UV has its main potential in the UV wavelength 
interval. The mission is an international collaboration, with 
Russia\footnote{\url{http://wso.inasan.ru/}}, 
Spain\footnote{\url{http://www.wso-uv.es/index.php/home.html}} and 
Germany\footnote{\url{http://www.uni-tuebingen.de/index.php?id=8711}} playing the 
leading role. 

WSO-UV is equipped with multipurpose instrumentation to carry out spectroscopy 
and imaging. In particular, we concentrate on the Channel for Surveys (CfS) of 
the Imaging and Slitless Spectrograph Instrument for Surveys (ISSIS). The ISSIS 
CCD (CfS) is characterized by
\begin{enumerate}
\item small pixels (0.05$\arcsec$), 
\item narrow distributions of brightness for point-like sources (FWHM = 
0.1$\arcsec$), and 
\item a field-of-view (FOV) = $3.4\arcmin \times 3.4\arcmin$. 
\end{enumerate}
Hence, the CfS is ideally suited for GLQ programmes, which require a high angular 
resolution (see Fig.~\ref{SelectionV20}) and a reasonably large FOV (to find 
reference stars). Its UV spectral coverage (1100-4000 \AA) corresponds to $\lambda 
\sim$ 440-1600 \AA\ in the rest-frame of bright GLQs at $z \sim$ 1.5, and thus, the 
CfS can be used to probe the EUV-FUV emission of GLQs. We propose to follow-up the 
variability of several targets over observation periods of about six months and 
using a dense sampling (e.g. two or three times per week). To optimise the 
monitoring of each GLQ, starting times (of observation periods) would be decided 
through an optical alert system (just after detections of prominent optical 
gradients). As we remark in Sect. 2, this time-domain programme has high potential 
for different cosmological studies. Moreover, very preliminary simulations (using 
the ISSIS Exposure Time Calculator, Alpha 2 version) suggest that it is not too 
time consuming. In Table~\ref{tbl-1}, we present some exposure times (broad-band 
filters) that are needed to obtain a signal-to-noise ratio (SNR) of 100 for a 
target with $V$ = 18 mag. Taking these (preliminary!) results into account, a 
whole campaign for a typical GLQ might last $\sim$ 645 s/d $\times$ 50 d $\sim$ 30 
ks (science time).

\begin{table}[t]
\caption{Exposure times required to obtain SNR = 100 ($V$ = 18 mag)\label{tbl-1}}
\centering
\begin{tabular}{@{}lcc}
\hline
Filter & Emission & Time (s) \\
\hline
F170W & EUV & $\sim$ 430 \\
F255W & EUV & $\sim$ 190 \\
F336W & FUV & $\sim$ 25  \\
\hline
\end{tabular}
\end{table}

\begin{table}[t]
\caption{Exposure times required to obtain SNR = 100 ($V$ = 17-20 mag)\label{tbl-2}}
\centering
\begin{tabular}{@{}lcc}
\hline
Filter & Time/$V$=17 (ks) & Time/$V$=20 (ks) \\
\hline
F170W & $\sim$ 0.17 & $\sim$ 2.63 \\
F255W & $\sim$ 0.08 & $\sim$ 1.19 \\
F280N & $\sim$ 0.40 & $\sim$ 6.14 \\
\hline
\end{tabular}
\end{table}

We also propose a second programme: {\it EUV database of GLQs}. Its main scientific 
goals are:
\begin{itemize}
\item To study nuclear and circumnuclear EUV emissions ($\lambda \leq$ 1200 \AA\ 
within a 0.3$\arcsec$ radius) of a slected sample of GLQs \citep[e.g.][]{hutc03}.

\item To compare with other observations (e.g. VLBI radio jets), and to contribute 
to a database covering the whole electromagnetic spectrum.

\item To reconstruct the EUV morphology of central regions in active galaxies, and 
to constraint gravitational lens scenarios.
\end{itemize}
Table~\ref{tbl-2} shows exposure times required to obtain SNR = 100 for targets with 
17 $\leq V \leq$ 20 mag. The faintest targets ($V$ = 20 mag) may only be reasonably
well imaged in a narrow filter from relatively long exposures (exceeding 6 ks). See
the GLENDAMA web site at \url{http://grupos.unican.es/glendama/index.htm} for 
updates on our GLQ project with WSO-UV. 

\acknowledgments
LJG is grateful to the organizers (LOC and SOC) of the international conference 
"Ultraviolet Universe 2010" held in St. Petersburg (Russia) between May 31 and June 3,
2010, who prepared a meeting showing UV facilities and studies, and including a long
list of complementary/social activities (the excursion along the Neva and Moika rivers
was really great!). I also thank all members of the Science Working Group (WSO-UV) in 
Spain for very interesting discussions. We use information taken from the CfA-Arizona 
Space Telescope LEns Survey (CASTLES) and SDSS Quasar Lens Search (SQLS) web sites, 
and we are grateful to both teams for doing that public databases. This research has 
been financially supported by the Spanish Department of Education and Science grant 
AYA2007-67342-C03-02 and University of Cantabria funds.

\end{document}